\documentclass[twocolumn,aps,amsmath,showpacs,superscriptaddress]{revtex4}
\usepackage{graphicx,dcolumn,bm,amssymb,amsmath,ulem,indentfirst,amsthm}
\usepackage[toc,page]{appendix}
\usepackage{color}

\theoremstyle{plain}
\def\be{\begin{equation}}
\def\ee{\end{equation}}

\newtheorem*{theorem*}{Theorem}

\begin{document}
\author{Bingyu Cui}
\affiliation{Cavendish Laboratory, University of Cambridge, JJ Thomson
Avenue, CB3 0HE Cambridge,
U.K.}
\author{Giancarlo Ruocco}
\affiliation{Department of Physics, University of Rome ''La Sapienza”, 00185 Roma, Italy}
\affiliation{Center for Life NanoScience, Istituto Italiano di Tecnologia, Roma,
Italy}
\author{Alessio Zaccone}
\email{az302@cam.ac.uk}
\affiliation{Cavendish Laboratory, University of Cambridge, JJ Thomson
Avenue, CB3 0HE Cambridge,
U.K.}
\affiliation{Department of Physics ``A. Pontremoli", University of Milan, via Celoria 16, 20133 Milano, Italy}
\affiliation{Statistical Physics Group, Department of Chemical
Engineering and Biotechnology, University of Cambridge, Philippa Fawcett Drive,
CB3 0AS Cambridge, U.K.}

\begin{abstract}
A new microscopic derivation of the elastic constants of amorphous solids is presented within the framework of nonaffine lattice dynamics, which makes use of a perturbative form of the low-frequency eigenvectors of the dynamical matrix introduced in [V. Mazzacurati, G. Ruocco, M. Sampoli EPL 34, 681 (1996)]. The theory correctly recovers the shear modulus at jamming, $\mu \sim (z-2d)$, including prefactors in quantitative agreement with simulations.
Furthermore, this framework allows us, for the first time, to include the effect of internal stresses. The theory shows that the Maxwell rigidity criterion $z=2d$ is violated with internal stress. In particular, 
$\mu \sim (z-2df)$ where $f<1$ if the bonds are, on average, stretched, and the solid is thus rigid below the Maxwell isostatic limit, while $f>1$ if the bonds are, on average, compressed. The coefficient $f$ is derived in analytical form and depends only on $d$ and on the average particle displacement from the interaction energy minimum. 
\end{abstract}

\title{Theory of elastic constants of athermal amorphous solids with internal stresses}
\maketitle

\section{Introduction}
In both thermal glasses and athermal granular solids, and unlike ordinary solids (e.g. crystals), the elasticity, mechanical stability and deformation behaviour are strongly affected by internal stresses, and by the local stress transmission, in the form of force-chains and force-contact networks~\cite{Behringer1,Behringer2,Behringer3}. 
Indeed, glasses can be described as emerging from the liquid through a mechanism of vitrification that can rationalize history-dependence and frozen stresses~\cite{Fuchs,Ballauf}. 

In order to elucidate the elasticity of amorphous solids in a unifying framework, the two most studied paradigms are given by random networks~\cite{Lubensky} (also used as a model for biological filaments such as the cytoskeleton, and other polymer-based materials) and random sphere packings~\cite{OHern} (a model for granular materials and jammed matter). 

While numerical simulations have substantially confirmed the picture emerging from experimental characterization of force transmission in granular and disordered materials, theory has somehow been left behind, with few exceptions~\cite{Blumenfeld}. The central problem is that it is very difficult to incorporate stress-transmission in theoretical models, such as lattice theories. As remarked in~\cite{Cates}, the reason is that ``In such a medium the displacement field is not single valued, and the solution of the elastic problem, though possible in principle, requires the whole construction history to be taken into account.'' The latter task is clearly challenging for theory.
As a consequence, all analytical theories of the elasticity and rigidity of amorphous solids proposed so far 
have specialized to the case of packings near the jamming point (where all forces vanish)~\cite{Mizuno}.

In this contribution, we attempt a step forward in the direction of incorporating internal stresses into the lattice dynamical theory of amorphous solids. The starting point, which allows us to proceed in this direction, is the formulation of an appropriate ansatz for the eigenvectors of amorphous solids which contains the effect of disorder as a random perturbation to wave-like components~\cite{Mazzacurati1996}. Using this form for the eigenvectors, we are able to evaluate the lattice dynamics for nonaffine deformations and we can consider the effect of internal stress in a mean-field way, in the two opposite limits of stretched bonds and compressed bonds.

\section{Formalism} 

The starting point of all lattice-dynamical theories is the Born-Huang free energy expansion \cite{Born}. The response is called affine if the interparticle displacements are just the old positions acted upon by the macroscopic strain tensor. The situation is different with a disordered or a non-centrosymmetric lattice where local inversion symmetry is absent~\cite{Milkus}. Consequently, forces that every particle receives from surrounding particles no longer cancel by symmetry, but they have to be relaxed with additional particle displacements because the whole system has to remain in mechanical equilibrium at every step in the deformation. The additional atomic displacements are known as nonaffine displacements. 

Throughout this paper, we closely follow the notation of Lemaitre and Maloney \cite{Lemaitre}. Roman indices are used to label particles while Greek indices label Cartesian components. Vectors in $d$-dimensional space is denoted by bold characters. For convenience, we assume that all particles have unit mass.

In the usual language of elasticity, particle positions in reference frame, $\mathring{\mathbf{r}}$, and current frame, $\mathbf{r}$, can be related by deformation gradient tensor $\underline{\underline{F}}$ in the way that $\mathbf{r}=\underline{\underline{F}}\mathring{\mathbf{r}}$. We further introduce the Cauchy-Green strain tensor $\underline{\underline{\eta}}=\frac{1}{2}(\underline{\underline{F}}^T\underline{\underline{F}}-\underline{\underline{I}})$ to describe the deformations, with which, the potential energy can be written either in reference frame or current frame, $\mathring{\mathcal{U}}(\{\mathbf{\mathring{r}}_i\},\underline{\underline{\eta}})\equiv\mathcal{U}(\{\mathbf{r}_i\},\underline{\underline{F}})$. $\underline{\underline{\eta}}$ is defined after the reference cell is chosen whereas functional $\mathcal{U}(\mathbf{r},\underline{\underline{F}})$ does not depend on this choice. Thus, for fixed $\{\mathring{\mathbf{r}}\}$, changing $\underline{\underline{\eta}}$ means the response to affine strain of the whole system; changing $\{\mathring{\mathbf{r}}\}$ in the reference configuration corresponds to additional non-affine displacement particles proceed. By differentiating the force $f_i^\alpha=-\frac{\partial\mathring{\mathcal{U}}}{\partial\mathring{r}_i^\alpha}$ atcing on particle $i$ with respect to the components of the strain tensor and taking limit $\underline{\underline{\eta}}\rightarrow0$, we obtain the equation of motions at mechanical equilibrium~\cite{Lemaitre,Zaccone2011}:
\begin{equation}
\sum_{j,\beta}H_{ij}^{\alpha\beta}\frac{\mathcal{D}\mathring{r}^\beta_j}{\mathcal{D}\eta_{\kappa\chi}}|_{\underline{\underline{\eta}}\rightarrow\underline{\underline{0}}}=\Xi_{i,\kappa\chi}^\alpha
\end{equation}
where the Hessian $H_{ij}^{\alpha\beta}$ and the affine force field $\Xi_{i,\kappa\chi}^\alpha$ are given respectively by
\begin{align}
H_{ij}^{\alpha\beta}&=\frac{\partial^2\mathring{\mathcal{U}}}{\partial\mathring{r}_i^\alpha\partial\mathring{r}_j^\beta}|_{\underline{\underline{\eta}}\rightarrow\underline{\underline{0}}}\notag\\
\Xi_{i,\kappa\chi}^\alpha&=-\frac{\partial^2\mathring{\mathcal{U}}}{\partial\mathring{r}_i^\alpha\partial\eta_{\kappa\chi}}|_{\underline{\underline{\eta}}\rightarrow\underline{\underline{0}}}.
\end{align}
$\mathcal{D}$ in Eq. (1) is the material or total derivative (see Ref. \cite{Lemaitre}).
The elastic constant tensor, which is given by second derivatives of the energy functional with respect to 
$\underline{\underline{\eta}}$ can be written as
\begin{align}
C_{\alpha\beta\kappa\chi}&\equiv\frac{1}{\mathring{V}}\frac{\mathcal{D}^2\mathring{\mathcal{U}}}{\mathcal{D}\eta_{\alpha\beta}\mathcal{D}\eta_{\kappa\chi}}|_{\underline{\underline{\eta}}\rightarrow\underline{\underline{0}}}\notag\\
&=\frac{1}{\mathring{V}}\left(\frac{\partial^2\mathring{\mathcal{U}}}{\partial\eta_{\alpha\beta}\partial\eta_{\kappa\chi}}+\frac{\partial^2\mathring{\mathcal{U}}}{\partial\mathring{\mathbf{r}}_i\partial\eta_{\alpha\beta}}\cdot\frac{\mathcal{D}\mathbf{\mathring{r}}_i}{\mathcal{D}\eta_{\kappa\chi}}\right)|_{\underline{\underline{\eta}}\rightarrow\underline{\underline{0}}}\notag\\
&=\frac{1}{\mathring{V}}\frac{\partial^2\mathring{\mathcal{U}}}{\partial\eta_{\alpha\beta}\partial\eta_{\kappa\chi}}|_{\underline{\underline{\eta}}\rightarrow\underline{\underline{0}}}-\sum_{ij}\sum_{\iota\xi}\frac{1}{\mathring{V}}\Xi_{i,\alpha\beta}^\iota(H_{ij}^{\iota\xi})^{-1}\Xi_{j,\kappa\chi}^\xi\notag\\
&=C^A_{\alpha\beta\kappa\chi}-C^{NA}_{\alpha\beta\kappa\chi}.
\end{align}
From this it is clear that the elastic constant is given by the affine (Born-Huang) elastic constant~\cite{Born} $C_{\alpha\beta\kappa\chi}^A$ with a \textit{negative} correction provided by the nonaffine term $C_{\alpha\beta\kappa\chi}^{NA}\geq0$ \cite{Lemaitre,Zaccone2011}.

\section{Approximation method}
Assuming a pairwise interaction between particles in contact, $\mathcal{U}(\{r_{ij}\})=\sum_{\langle ij\rangle}V_{ij}(r_{ij})$, where the sum runs over all pairs of particles in contact $\langle ij\rangle$. We denote tension and stiffness (spring constant), respectively, as
\begin{equation}
t_{ij}=\frac{\partial V_{ij}}{\partial r_{ij}},~~s_{ij}=\frac{\partial^2 V_{ij}}{\partial r_{ij}^2}
\end{equation}
in terms of which, the affine force field and Hessian (dynamical) matrix are expressed as 
\begin{align}
\Xi_{i,\kappa\chi}^\alpha&=-\sum_j(r_{ij}s_{ij}-t_{ij})n_{ij}^{\alpha}n_{ij}^{\kappa}n_{ij}^{\chi}\\
H_{ij}^{\alpha\beta}&=\begin{cases}
-(s_{ij}-\frac{t_{ij}}{r_{ij}})n_{ij}^\alpha n_{ij}^\beta-\frac{t_{ij}}{r_{ij}}\delta_{\alpha\beta},\quad i\neq j\\
\sum_{k\neq i}(s_{ik}-\frac{t_{ik}}{r_{ik}})n_{ik}^\alpha n_{ik}^\beta+\frac{t_{ik}}{r_{ik}}\delta_{\alpha\beta},\quad i=j.
\end{cases}
\end{align}
We have used the identity $\partial/\partial\mathbf{r}_{ij}=\underline{n}_{ij}\partial/\partial r_{ij}$, with $\mathbf{n}_{ij}=\mathbf{r}_{ij}/r_{ij}$. 
Note that in a system with $N$ particles in total, and in $d$ space dimensions, the Hessian is a $dN\times dN$ symmetric semipositive-definite matrix with $d$ zero eigenvalues due to translational invariance (trivial Goldstone modes). Hence the product $dN$ denotes the total number of degrees of freedom in the system.
The affine and nonaffine part of elastic constant might be written as (the ring is dropped)
\begin{align}
C_{\kappa\chi\iota\xi}^{A}&=\frac{1}{V}\sum_{<ij>}(r_{ij}s_{ij}-t_{ij})r_{ij}n_{ij}^\kappa n_{ij}^\chi n_{ij}^\iota n_{ij}^\xi\notag\\
C_{\kappa\chi\iota\xi}^{NA}&=\frac{1}{V}\sum_{s=1, \lambda(s)\neq0}^{dN}\frac{(\underline{\Xi}_{\kappa\chi},\underline{\mathbf{v}}(s))(\underline{\Xi}_{\iota\xi},\underline{\mathbf{v}}(s))}{\lambda(s)},
\end{align}
where in $C^{NA}$ we implement normal mode decomposition and $\underline{\mathbf{v}}(s)$ (or $v_{i,\alpha}(s)$) are the eigenvectors of the Hessian, while $\lambda$ denotes the corresponding eigenvalues and 
$(,)$ means the normal scalar product on $\mathcal{R}^{dN}$.

In an approximation (supported by simulations) suggested in \cite{Mazzacurati1996}, one can model the (normalized) eigenvectors as sinusoidal waves with wave number $k_s=\omega_s/v$ plus a random component, $\epsilon_i(s)$, with zero average, and with variance $\sigma^2=\langle\epsilon_i^2(s)\rangle$ independent of normal mode $s\in\{1,2,...,dN\}$, i.e.
\begin{equation}
v_{i,\alpha}(s)=\hat{e}_\alpha\frac{1}{\sqrt{dN}}\left[\sqrt{2(1-\sigma^2)}\sin{(\mathbf{k}_s\cdot\mathbf{r}_i)}+\epsilon_i(s)\right]
\end{equation}
where $\hat{e}_\alpha,\alpha=x,y,z$ is the polarization unit vector such that $\hat{e}_\alpha\hat{e}_\beta=\delta_{\alpha\beta}$.

We define the angular average as:
\begin{equation}
\sum_{i=1}^N\sin^n(\mathbf{k}_s\cdot \mathbf{r}_i)\epsilon_{i}^m(s)=N\langle\sin^n(\mathbf{k_s}\cdot \mathbf{r}_i)\rangle\langle\epsilon_{i}^m(s)\rangle
\end{equation}
where $m,n$ are integers. 

For the case $n=2$, which is of interest for the normalization of the eigenvectors, the average can be evaluated as follows. Assuming translational invariance (which is justified for a uniform amorphous system at least in the low-$k$ limit), there is complete freedom in choosing or shifting the origin of the reference frame, i.e. $\mathbf{r} \rightarrow \mathbf{r} + \mathbf{r}_{0}$, where $\mathbf{r}_{0}$ is an arbitrary shift.   

Hence, 
$\langle \sin^2(\mathbf{k} \cdot \mathbf{r})\rangle = \langle \sin^2(\mathbf{k} \cdot (\mathbf{r} + \mathbf{r}_{0}))\rangle
= \langle \sin^2(\mathbf{k} \cdot \mathbf{r} + \mathbf{k} \cdot \mathbf{r}_{0})\rangle$. 
Next we define $\alpha \equiv \mathbf{k}
\cdot \mathbf{r}_{0}$, from which we get $\langle \sin^2(\mathbf{k} \cdot \mathbf{r}) \rangle = \langle \sin^2(\mathbf{k} \cdot \mathbf{r} + \alpha) \rangle$. 
Since $\alpha$ is an arbitrary scalar, we can choose $\alpha = \pi/2$, without loss of generality, since the identity must hold for any values of $\alpha$. Then, we see that $\langle \sin^2(\mathbf{k} \cdot \mathbf{r}) \rangle = \langle \cos^2(\mathbf{k} \cdot \mathbf{r}) \rangle$, which implies 
\begin{equation}
\langle \sin^2(\mathbf{k} \cdot \mathbf{r}) \rangle = \frac{1}{2}.
\end{equation}

Using this result, one can easily check that $v_{i,\alpha}$ is normalized,
\begin{align}
\sum_i\sum_\alpha v_{i,\alpha}^2&=\sum_i\frac{d}{Nd}\left[{2(1-\sigma^2)\sin^2{(\mathbf{k}_s\cdot \mathbf{r}_i)}}\right.\notag\\
&\left.{+2\sqrt{2(1-\sigma^2)}\sin(\mathbf{k}_s\cdot \mathbf{r}_i)\epsilon_i+\epsilon_i^2}\right]\notag\\
&=\frac{1}{N}\left[2(1-\sigma^2)\cdot\frac{N}{2}+N\sigma^2\right]=1\notag.
\end{align}
We want to find the form of an eigenvalue $\lambda$ of an eigenvector $\underline{\mathbf{v}}$, i.e. $H\mathbf{\underline{v}}=\lambda\mathbf{\underline{v}}$, that is
\begin{align}
&\left[H\mathbf{\underline{v}}\right]_{i,\alpha}=\sum_{j,\beta}H_{ij}^{\alpha\beta}v_{j,\beta}=\sum_{j\neq i}\sum_{\beta}H_{ij}^{\alpha\beta}v_{j,\beta}+\sum_{\beta}H_{ii}^{\alpha\beta}v_{i,\beta}\notag\\
&=\sum_{j\neq i}\sum_{\beta}\left[(s_{ij}-\frac{t_{ij}}{r_{ij}})n_{ij}^\alpha n_{ij}^\beta+\frac{t_{ij}}{r_{ij}}\delta_{\alpha\beta}\right](v_{i,\beta}-v_{j,\beta})\notag\\
&=\sum_{j\neq i}\sum_{\beta}\left(s_{ij}-\frac{t_{ij}}{r_{ij}}\right)n_{ij}^\alpha n_{ij}^\beta(v_{i,\beta}-v_{j,\beta})\notag\\
&+\sum_{j\neq i}\frac{t_{ij}}{r_{ij}}(v_{i,\alpha}-v_{j,\alpha})\notag\\
&=\sum_{j\neq i}\frac{1}{d}(s_{ij}-\frac{t_{ij}}{r_{ij}}+d\frac{t_{ij}}{r_{ij}})(v_{i,\alpha}-v_{j,\alpha})\notag\\
&=\frac{1}{d}\sum_{j\neq i}\left[s_{ij}+(d-1)\frac{t_{ij}}{r_{ij}}\right]v_{i,\alpha}\notag\\
&\equiv\lambda v_{i,\alpha}
\end{align}
where on the 3rd line, the orientational-dependent factors $n_{ij}^{\alpha}n_{ij}^{\beta}$ for a large system with uncorrelated isotropic disorder has been replaced with its isotropic (angular) average, i.e. $n_{ij}^{\alpha}n_{ij}^{\beta}\rightarrow\delta_{\alpha\beta}/d$.

With these approximations, we are able to obtain $(\underline{\Xi}_{\kappa\chi},\underline{\mathbf{v}}(s))(\underline{\Xi}_{\iota\xi},\underline{\mathbf{v}}(s))$ in analytical form:
\begin{widetext}
\begin{align}
&(\underline{\Xi}_{\kappa\chi},\underline{\mathbf{v}}(s))(\underline{\Xi}_{\iota\xi},\underline{\mathbf{v}}(s))
=\sum_{i,i'}^N\sum_{\alpha,\beta}\Xi_{i,\kappa\chi}^{\alpha}v_i^{\alpha}\Xi_{i',\iota\xi}^\beta v_{i'}^\beta=\sum_{i,i',j,j'}^N\sum_{\alpha,\beta}^d(r_{ij}s_{ij}-t_{ij})(r_{i'j'}s_{i'j'-t_{i'j'}})n_{ij}^\alpha n_{ij}^\kappa n_{ij}^\chi n_{i'j'}^\beta n_{i'j'}^\iota n_{i'j'}^\xi\notag\\
&\times\frac{1}{dN}\hat{e}^\alpha\hat{e}^\beta\left[\sqrt{2(1-\sigma^2)}\sin(\mathbf{k}_s\cdot\mathbf{r}_i)+\epsilon_i\right]
\left[\sqrt{2(1-\sigma^2)}\sin(\mathbf{k}_s\cdot\mathbf{r}_{i'})+\epsilon_{i'}\right]\notag\\
&=\frac{1}{dN}\sum_\alpha^d B_{\alpha,\kappa\chi\iota\xi}\sum_{i,i',j,j'}(\delta_{ii'}\delta_{jj'}-\delta_{ij'}\delta_{i'j})(r_{ij}s_{ij}-t_{ij})(r_{i'j'}s_{i'j'-t_{i'j'}})\notag\\
&\times\left[2(1-\sigma^2)\sin(\mathbf{k}_s\cdot\mathbf{r}_i)\sin(\mathbf{k}_s\cdot\mathbf{r}_{i'})+\epsilon_i\sqrt{2(1-\sigma^2)}\sin(\mathbf{k}_s\cdot\mathbf{r}_{i'})
+\epsilon_{i'}\sqrt{2(1-\sigma^2)}\sin(\mathbf{k}_s\cdot\mathbf{r}_{i})+\epsilon_{i}\epsilon_{i'}\right]\notag\\
&=\frac{1}{dN}\sum_\alpha^d B_{\alpha,\kappa\chi\iota\xi}\sum_{i,j}^N(r_{ij}s_{ij}-t_{ij})^2\left[{2(1-\sigma^2)\sin^2(\mathbf{k}_s\cdot\mathbf{r}_i)
+2\epsilon_i\sqrt{2(1-\sigma^2)}\sin(\mathbf{k}_s\cdot\mathbf{r}_{i'})+\epsilon_{i}^2}\right.\notag\\
&\left.{-2(1-\sigma^2)\sin(\mathbf{k}_s\cdot\mathbf{r}_i)\sin(\mathbf{k}_s\cdot\mathbf{r}_j)
-\epsilon_{i}\sqrt{2(1-\sigma^2)}\sin(\mathbf{k}_s\cdot\mathbf{r}_{j})-\epsilon_{j}\sqrt{2(1-\sigma^2)}\sin(\mathbf{k}_s\cdot\mathbf{r}_{i})-\epsilon_i\epsilon_j}\right]\notag\\
&=\frac{z}{d}\sum_\alpha^d B_{\alpha,\kappa\chi\iota\xi}\langle(r_{ij}s_{ij}-t_{ij})^2\rangle.
\end{align}
\end{widetext}
Here, upon taking an isotropic average, the term $n_{ij}^\alpha n_{ij}^\kappa
n_{ij}^\chi n_{i'j'}^\alpha n_{i'j'}^\iota n_{i'j'}^\xi$ may be replaced
with $(\delta_{ii'}\delta_{jj'}-\delta_{ij'}\delta_{i'j})B_{\alpha,\kappa\chi\iota\xi}$.
The $B_{\alpha,\kappa\chi\iota\xi}$ are geometric coefficients resulting from the angular average and are tabulated in~\cite{Zaccone2011}. In the last line, $z$ is the coordinate number labeling the number of nearest neighborhoods of a tagged particle. Assuming disorder is spatially uncorrelated, the angular and radial average have been taken separately.

\subsection{Unstressed random network}
We first consider the case of a random network of central-force springs, with no internal stress.  The interaction potential is a harmonic potential $V(r_{ij})=\frac{\kappa}{2}(r_{ij}-R_0^2)$. Also, $\kappa \equiv s_{ij}$ is the spring constant and $R_0$ is the distance between two particles in contact in the reference frame. The reference state is unstressed, i.e, all springs are relaxed in the minimum of the harmonic well. Hence, $t_{ij}\equiv 0$ and $r_{ij}\equiv R_0$.

For the nonaffine part of the elastic stiffness tensor, we have
\begin{align}
C_{\kappa\chi\iota\xi}^{NA}&=\frac{1}{V}\cdot dN\cdot \frac{zR_0^2\kappa^2\sum_{\alpha}B_{\alpha,\kappa\chi\iota\xi}}{d\cdot\frac{1}{d}z\kappa}\notag\\
&=\frac{dNR_0^2\kappa}{V}\sum_{\alpha}B_{\alpha,\kappa\chi\iota\xi}.
\end{align}
The eigenvalue $\lambda$ is actually independent of normal mode label $s$. Likewise, the affine (Born) term can be expressed as 
\begin{align}
C_{\kappa\chi\iota\xi}^{A}&=\frac{NzR_0^2\kappa}{2V}\langle n_{ij}^\kappa n_{ij}^\chi n_{ij}^\iota n_{ij}^\xi\rangle.
\end{align}
Here $\langle n_{ij}^\kappa n_{ij}^\chi n_{ij}^\iota n_{ij}^\xi\rangle=\sum_{\alpha}^dB_{\alpha,\kappa\chi\iota\xi}$, and we write the elastic constant tensor as
\begin{equation}
C_{\kappa\chi\iota\xi}=C_{\kappa\chi\iota\xi}^{A}-C_{\kappa\chi\iota\xi}^{NA}=\frac{NR_0^2\kappa}{2V}\sum_{\alpha=1}^dB_{\alpha,\kappa\chi\iota\xi}(z-2d)
\label{elastic:cauchy}
\end{equation}
For the shear modulus, $\sum_{\alpha}^d B_{\alpha,xyxy}=1/15$ and Eq. (\ref{elastic:cauchy}) recovers the same analytical results of \cite{Zaccone2011}, without any fitting parameters. The prefactor has been compared with the simulations in $d=3$ of random frictionless packings near jamming interacting via harmonic potential in \cite{OHern}, and an excellent quantitative agreement was found even for the prefactor $\frac{NR_0^2\kappa}{30V}$. 

One should note that the prediction for the bulk modulus does not describe what is found in jammed packings, where $K \sim z$, instead of $K \sim (z-2d)$. The reason was explained in \cite{Ellenbroek} and has to do with the short-range particle correlations in the packing, which alter the affine force balance on the particle and reduce the nonaffinity. Upon duly accounting for these correlations, the correct scaling can be recovered within the present framework, as shown in \cite{Zaccone2014,Zaccone2016}.
 
\subsection{Random network with internal stresses}
Now we weaken the condition that the interparticle distance $R_0$ coincides with the minimum of the harmonic potential, but we introduce a distribution of interparticle distances peaked at an average value $R_{e}\neq R_{0}$. On average, we let $r_{ij}\equiv R_e$.
The fact that the actual distance between two particles in contact deviates from the minimum of the interaction automatically implies the existence of a bond-tension or stress. In other words, the spring is either compressed, $R_{e}<R_{0}$, or stretched, $R_{e}>R_{0}$.

It was pointed out by S. Alexander with the famous metaphor of the violin strings, that internal stresses, which cause bonds to stretch, can make underconstrained lattices (i.e. with $z<2d$) fully rigid, which otherwise would be floppy~\cite{Alexander}. 
From numerical simulations it is also known that, in disordered elastic networks, internal stresses have a profound effect on mechanical response, and can indeed make underconstrained lattices rigid ~\cite{Huisman}. 

With these model assumptions we then get 
\begin{align}
C^{A}_{\kappa\chi\iota\xi}&=\frac{NzR_0R_e\kappa}{2V}\langle n_{ij}^\kappa n_{ij}^\chi n_{ij}^\iota n_{ij}^\xi\rangle\notag\\
&=\frac{NzR_0R_e\kappa}{2V}\sum_\alpha^d B_{\alpha,\kappa\chi\iota\xi}
\end{align}
To evaluate the nonaffine contribution to elastic constants $C^{NA}_{\kappa\chi\iota\xi}$, we get
\begin{subequations}
\begin{align}
&(\underline{\Xi}_{\kappa\chi},\underline{\mathbf{v}}(s))(\underline{\Xi}_{\iota\xi},\underline{\mathbf{v}}(s))\\
=&\frac{1}{dN}\sum_\alpha^d B_{\alpha,\kappa\chi\iota\xi}\cdot Nz\cdot (\kappa R_0)^2=\frac{z \kappa^2R_0^2}{d}\sum_\alpha^d B_{\alpha,\kappa\chi\iota\xi}\\
\lambda&=\frac{1}{d}\sum_{j\neq i}\left[s_{ij}+(d-1)\frac{t_{ij}}{r_{ij}}\right]\notag\\
&=\frac{z}{d}\left[\kappa+\kappa(d-1)(1-\frac{R_0}{R_e})\right]\notag\\
C_{\kappa\chi\iota\xi}^{NA}&=\frac{1}{V}\cdot dN\cdot \frac{dz\kappa^2R_0^2}{dz\left[\kappa+\kappa(d-1)(1-\frac{R_0}{R_e})\right]}\sum_\alpha^d B_{\alpha,\kappa\chi\iota\xi}\notag\\
&=\frac{dN\kappa R_0^2}{V\left[1+(d-1)(1-\frac{R_0}{R_e})\right]}\sum_\alpha^d B_{\alpha,\kappa\chi\iota\xi},
\end{align}
\end{subequations}
where we used $t_{ij}=\kappa(R_e-R_0)$.
Finally, the elastic constants for the network with internal stress may be expressed as
\begin{align}
C_{\kappa\chi\iota\xi}&=C_{\kappa\chi\iota\xi}^{A}-C_{\kappa\chi\iota\xi}^{NA}\notag\\
&=\frac{N\kappa R_0R_e}{2V}\left[z-\frac{2d\frac{R_0}{R_e}}{1+(d-1)(1-\frac{R_0}{R_e})}\right]\sum_\alpha^d B_{\alpha,\kappa\chi\iota\xi}
\end{align}
Obviously, if $R_e=R_0$, then we recover results in Section III.A. 

To summarize, we have found that with internal stresses, the elastic constants (including the shear modulus $C_{xyxy}\equiv\mu$) are given by:
\begin{align}
&C_{\kappa\chi\iota\xi}\sim(z-2df),\\
&f=\frac{R_0/R_e}{1+(d-1)(1-\frac{R_0}{R_e})}.
\end{align}
If $R_e<R_0$, then $f>1$; if $R_e<R_0$, then $f<1$. Fig. 1 shows, when $d=2$, how the ratio $R_e/R_0$ affects the dependence of $C_{\kappa\chi\iota\xi}$ on $z$.\\

\begin{figure}[h]
\centering
\includegraphics[height=2in]{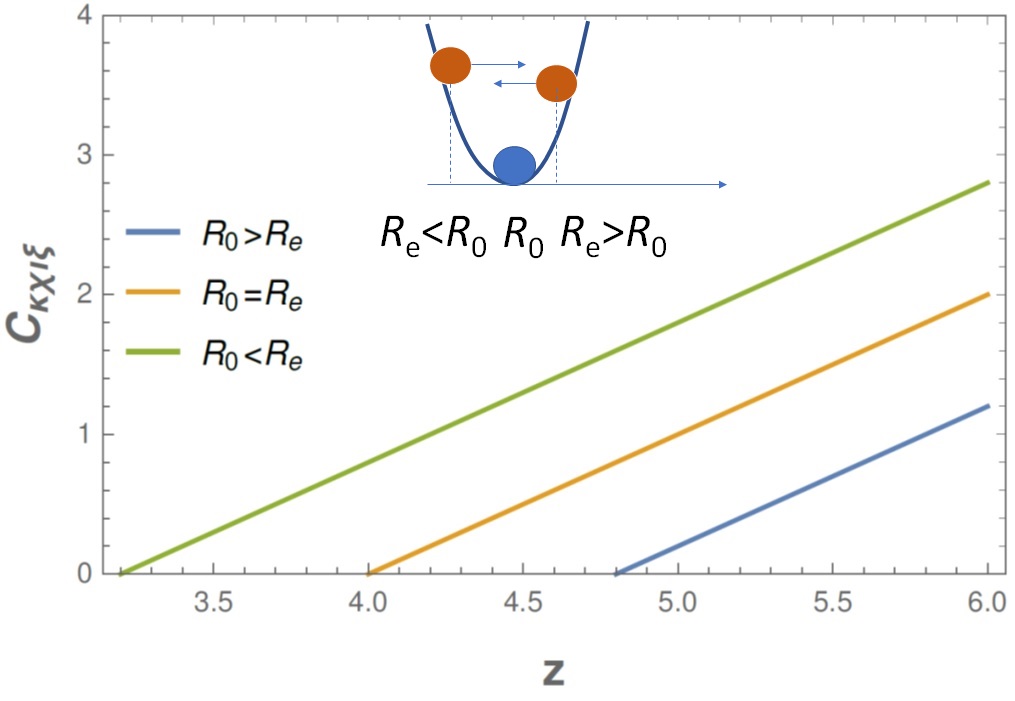}
\caption{Sketch of the dependence of the elastic constant $C_{\kappa\chi\iota\xi}$ as a function of coordination $z$ for different values of the internal stress parameter $R_{e}/R_{0}$ which indicates the initial particle displacement from the interaction minimum. Results are obtained in the 2-dimensional system.}
\end{figure}

From a physical point of view, the behavior seen in Fig. 1, means that when the internal strain is raised due to initially stretched network bonds, then larger elastic constants are required to "pull back" particles to equilibrium positions. On the other hand, if the bonds are initially compressed, the elastic constants become smaller. The latter effect is more unexpected and calls for further verifications using numerical simulations, although it may already have been detected in simulations of colloidal gels upon preparing particles under conditions of compression of bonded neighbours~\cite{DelGado}.
The fact that pre-stretched bonds lead to a larger elastic modulus confirms an earlier intuition of S. Alexander~\cite{Alexander}. 

\section{Conclusion}
We presented a fully analytical derivation of the elastic constants of athermal disordered solids within the framework of nonaffine lattice dynamics. A perturbative ansatz for the eigenvectors of the dynamical (Hessian) matrix has been used to evaluate the nonaffine contribution to the elastic constants. For random assemblies of harmonically interacting particles, in the limit of particles being in the minimum of the harmonic interaction the theory recovers well known results and the scaling $\mu \sim (z-2d)$ for the shear modulus, including the prefactor in quantitative agreement with simulations.
When the particles are initially away from the minimum, initial stresses are present. Two opposite limits of bonds being on average stretched and bonds being on average compressed have been considered. For pre-stretched bonds the system is rigid also below the Maxwell rigidity threshold, whereas for pre-compressed bonds the onset of rigidity is shifted to higher coordination number. This might be a first step in the direction of a theory of granular matter and disordered solids where internal stresses are explicitly taken into account.

\section{Ethical statement}
Conflict of Interest: The authors declare that they have no conflict of interest.\\
Novelty of the content of the paper: The paper is not currently being considered for publication by other journals and was not previously published. All the data is authentic and was developed by the authors.\\
Funding:  As  stated  in  the  acknowledgements  section, the  Ph.D.  studies  of  the  first  author  are  funded  by the CSC-Cambridge Scholarship.

\begin{acknowledgements}
B. C acknowledges the financial support of CSC-Cambridge Scholarship.  Eugene M. Terentjev and Fanlong Meng are gratefully acknowledged for providing input and many useful discussions. 
\end{acknowledgements}

\end{document}